\documentclass[12pt]{article}
\usepackage{scicite}
\usepackage{times}
\usepackage{graphicx}
\usepackage{setspace}
\usepackage{color}
\usepackage{titling}

\newenvironment{sciabstract}{%
\begin{quote} \bf}
{\end{quote}}

\title{Zeptosecond Birth Time Delay in Molecular Photoionization} 

\author
{Sven Grundmann,$^{1\ast}$ Daniel Trabert,$^{1}$ Kilian Fehre,$^{1}$ Nico Strenger,$^{1}$ \\
 Andreas Pier,$^{1}$ Leon Kaiser,$^{1}$ Max Kircher,$^{1}$ Miriam Weller,$^{1}$\\
  Sebastian Eckart,$^{1}$ Lothar Ph. H. Schmidt,$^{1}$ Florian Trinter,$^{1,2,3}$\\
   Till Jahnke,$^{1\dagger}$ Markus S. Sch\"offler,$^{1}$
 Reinhard D\"orner$^{1\ddagger}$\\
\\
\normalsize{$^{1}$Institut f\"ur Kernphysik, Goethe-Universit\"at,}\\
\normalsize{Max-von-Laue-Strasse 1, 60438 Frankfurt, Germany}\\
\normalsize{$^{2}$Photon Science, Deutsches Elektronen-Synchrotron (DESY),}\\
\normalsize{Notkestrasse 85, 22607 Hamburg, Germany}\\
\normalsize{$^{3}$Molecular Physics, Fritz-Haber-Institut der Max-Planck-Gesellschaft,}\\
\normalsize{Faradayweg 4-6, 14195 Berlin, Germany}\\
\\
\normalsize{$^\ast$  grundmann@atom.uni-frankfurt.de}\\
\normalsize{$^\dagger$  jahnke@atom.uni-frankfurt.de}\\
\normalsize{$^\ddagger$  doerner@atom.uni-frankfurt.de}
}

\date{}

\begin{document} 


\baselineskip24pt


\pretitle{%
  
  \begin{singlespace}
  \textcolor{red}{
This is the author’s version of the work.
The definitive version was published in \textit{Science}:\\
16 Oct 2020, Vol. 370, Issue 6514, pp. 339-341, DOI: 10.1126/science.abb9318.  
  }   
  \end{singlespace}
  \begin{center}
  \LARGE
}
\posttitle{\end{center}}

\maketitle

One Sentence Summary:
\begin{sciabstract}
We observe zeptosecond delays in photoelectron emission corresponding to the travel time of the light along a molecular orbital.
\end{sciabstract}

\newpage
Abstract:
\begin{sciabstract}
Photoionization is one of the fundamental light-matter interaction processes in which the absorption of a photon launches the escape of an electron.
The time scale of the process poses many open questions.
Experiments found time delays in the attosecond ($10^{-18}$ s) domain between electron ejection from different orbitals, electronic bands, or in different directions.
Here, we demonstrate that across a molecular orbital the electron is not launched at the same time.
The birth time rather depends on the travel time of the photon across the molecule, which is 247 zeptoseconds ($10^{-21}$ s) for the average bond length of H$_2$. 
Using an electron interferometric technique, we resolve this birth time delay between electron emission from the two centers of the hydrogen molecule.
\end{sciabstract}

\newpage
\section*{Main Text}
Photoionization is a fundamental quantum process and has become a powerful tool to study atoms, molecules, liquids, and solids.
Facilitated by the advent of attosecond technology, it can nowadays even be addressed in the time domain.
Timing in photoionization usually refers to the time it takes an electron to escape to the continuum after absorption of the photon.
This time depends, e.g., on the electronic orbital \cite{Schultze2010,Isinger2017},
on the energy band in solids~\cite{Cavalieri2007,Tao2016},
or on the orientation \cite{Vos2018} and handedness \cite{Beaulieu2017} of the target molecule.
The escape time difference manifests as a phase shift of the electron wave in the far field.
This relation is based on the concept of the Wigner delay, which is the energy derivative of the phase of a photoelectron wave at an asymptotic distance from the source \cite{Dahlstrom2012}.
Typical numbers are, e.g., 20 attoseconds (as = $10^{-18}$~s) for the Wigner delay between emission from the $2s$ and $2p$ shells in neon \cite{Schultze2010}.

The Wigner delay, however, does not cover another intriguing question on timing in photoionzation of extended systems,
namely the temporal buildup of the photoelectron wave across its spatially extendend source.
In the following, we refer to these variations in the temporal structure of the electron wave as the birth time delay $\tau_b$.
Although the Wigner delay is caused during the travel of the electron to the continuum after its birth, $\tau_b$ relies on different birth times of the contributions to the total photoelectron wave along a molecular orbital.
Accordingly, the birth time delay quantifies to which extend a delocalized molecular orbital reacts simultaneously as one single unit upon being hit by a photon.
For example, it shows if the part of the orbital facing towards an approaching photon reacts first and the part downstream of the photon beam has a retarded response.
Interestingly, the delays expected from that travel time of a photon across molecular orbitals are one to two orders of magnitude shorter than the Wigner delay, i.e., they manifest in the zeptosecond (zs = $10^{-21}$~s) domain. 
In the description of light-matter interaction, such ultrashort time differences are often ignored and the dipole approximation is invoked.  
It corresponds to neglecting the spatial dependence of the light wave,  
and the light's electromagnetic field is approximated to be present instantaneously with the same phase over the whole relevant region of space.
Beyond this approximation, however, the birth time delay leads to a phase shift between the corresponding contributions to the overall photoelectron wave.
Such relative phases -- and hence the birth time delays across a molecular orbital~-- are accessible by experiments exploiting the interference between the different parts of the wavefunction.
Figure 1 outlines our metrology to measure $\tau_b$.

Our approach builds on the close analogy between a plane wave behind a double slit (Fig. 1, A and B) and the photoelectron wave emitted from a homonuclear diatomic molecule (Fig. 1, C and D).
This analogy was proposed by Cohen and Fano \cite{Cohen1966} and is well established today \cite{Fernandez2007,Canton2011}. 
It has been used, e.g., to study the onset of decoherence of a quantum system \cite{Akoury2007} and entanglement in an electron pair \cite{Waitz2016}. 
The angular emission pattern from a gerade orbital has a maximum perpendicular to the molecular axis, which corresponds to the zeroth-order interference maximum behind the double slit.
If a constant phase shift $\Delta \phi$ is introduced to one of the slits, the interference pattern behind the double slit becomes asymmetric, and the angular position of the interference fringes moves as function of $\Delta \phi$ on an intensity screen in the far field.
In Fig. 1, A and B we illustrate this relation.
A corresponding shift may also occur when two interfering electron waves are emerging from the two indistinguishable centers of a homonuclear diatomic molecule upon photonabsorption (Fig. 1, C and D) .
If the contributions to the photoelectron wave are launched simultaneously across an orbital of a diatomic molecule, the electron emission pattern in the molecular frame of reference is symmetric with respect to the normal of the bond axis.
However, any initial phase shift between the waves emerging from one or the other center leads to an angular shift of the diffraction pattern -- just as in the double-slit case.
Hence, inspecting the angular emission distribution of photoelectrons emitted from a homonuclear diatomic molecule for such angular shifts offers a way to measure the birth time delay $\tau_b$ \cite{Note1}.

For an electron of kinetic energy $E_e$ and momentum $p_e$, $\tau_b$ can be obtained from the measured angle $\alpha_0$, to which the central interference maximum is shifted, in the following way:
An electron wave is emitted and propagates with the phase velocity $v_{ph}=E_e / p_e$  for the time $\tau_b$ before a second electron wave is born at a distance $R$.
The zeroth-order interference maximum occurs under the emission angle $\alpha_0$ for which the path length difference between both electron waves vanishes:
\begin{equation}
2 \pi\left( \frac{\cos(\alpha_0) \cdot R}{\lambda} - \frac{\tau_b \cdot v_{ph}}{\lambda} \right) = 0 ~,
\end{equation}
where $\lambda$ is the electron's de~Broglie wavelength.
Accordingly, the birth time delay $\tau_b$ can be inferred from the angular shift $\alpha_0$, and an angle of $\alpha_0=90^\circ$ corresponds to zero birth time delay (simultaneous emission):
\begin{equation}
\tau_b = \cos(\alpha_0) \frac {R}{v_{ph}}~. \label{equntau}\\
\end{equation}

We implemented this scheme by studying one-photon double ionization of H$_2$ using left-handed circularly polarized photons with an energy of 800~eV.
Using a COLTRIMS reaction microscope \cite{Ullrich2003}, 
we measured the three-dimensional momenta of both protons in coincidence with one electron.
From the sum momentum of the three measured particles we inferred the missing electron momentum vector via momentum conservation including the photon momentum.
The molecules in the target gas jet were randomly oriented with respect to the light propagation.
After ejection of the two electrons, the two protons were driven apart by their Coulomb repulsion.
We obtained the orientation of the molecular axis and the kinetic energy release (KER) from the relative momentum of the protons \cite{Weber2004,Schoffler2008a}.
In the reflection approximation, the internuclear distance $R$ at the moment of photoabsorption is related to the kinetic energy release via $KER = e^2/(4 \pi \epsilon_0 R)$, where $e$ is the elementary charge and $\epsilon_0$ is the vacuum permittivity.

One-photon double ionization typically proceeds via one of two sequential processes.
A primary photoelectron is set free by the absorption of the photon and the second electron is either shaken off or knocked out to the continuum \cite{Siedschlag2005}. 
The fraction of the photon energy that exceeds the sum of the adiabatic double ionization energy of H$_2$ (31.03~eV) and the KER is shared between the two electrons.
The symmetry of this energy sharing is a measure for the strength of the Coulomb interaction among the two electrons which has the potential to destroy the single-particle quantum interference pattern \cite{Waitz2016}.
Therefore, we restricted our investigation to fast electrons that carried more than 96\% of the excess energy (Fig. S2).
For such fast electrons, double-slit interference effects are well established on the single-particle level \cite{Akoury2007,Horner2008}.
Corresponding slow electrons of the double ionization process possessed less than 4\% of the excess energy and are not shown here \cite{Note2}.

In Fig. 2 A, we display the electron angular distribution of those fast electrons for the average internuclear distance of 0.74~\AA ~(purple).
The results show a rich structure, which --~as expected~-- resembles the interference pattern of electrons emerging from a double slit. 
Figure 2 B displays this measured interference pattern of the fast electron as function of the internuclear distance.
The results show how the number of interference fringes increased and how the angular separation of maxima decreased with increasing internuclear distance $R$, affirming the double-slit nature of the electron emission.

The data shown in Fig. 2 were averaged over all orientations of the molecular axis with respect to the light propagation and the light’s polarization plane.
Thus, the results must be symmetric.
To search for possible shifts of the interference fringes due to birth time delays, we inspected the interference pattern of the fast electron for different angles $\beta$ between the photon propagation direction and the molecular axis in Fig. 3 for the subset $\mathcal{S}$ (see Fig. 2 B).
Figure~3~A shows the measured fringes for molecules aligned parallel to the light propagation direction (see Fig. S3 for a corresponding polar plot).
The zeroth-order maximum of the distribution was displaced to the right which suggested the existence of a birth time delay.
To confirm this assumption, we depicted the interference fringes as function of $\cos(\beta)$ in Fig. 3 B.
The histogram shows a clear dependence of the central fringe on the photon direction.
For a quantitative analysis we determined $\cos(\alpha_0)$, i.e., the angular position of the central maximum, for each row of the histogram in Fig. 3 B via a Gaussian fit (red curve in Fig. 3 A).
Figure 3 C shows the results of these fits and the corresponding birth time delays (according to Eq. \ref{equntau}) on the right vertical scale.
Note that the birth time delay might be interpreted as a nondipole Wigner delay between photoionization from different locations of the spatially extended molecular orbital.
However, the usual Wigner times -- treated entirely within the dipole approximation so far -- would likely depend on $\alpha$ and $\beta$, but they are equal for the different pathways that interfere under a certain emission angle $\alpha$ and do not influence our measurement.

We compared our experimental findings to two simple models.
First, we assumed that $\tau_b'$ is given by the time difference with which a point of constant phase of the photon wave hits the two centers, 
\begin{equation}
\tau_b' = \cos(\beta) \frac{R}{c}~,
\end{equation}
where $c$ is the speed of light.
In the case of $\cos(\beta)=0$, the photon wave hits both centers of the molecule simultaneously and there cannot be any birth time delay between electrons emitted from one or the other center because the outgoing waves are exactly in phase.
On the other hand, $\cos(\beta)=\pm 1$ resembles the maximum possible travel time of the photon from one molecular center to the other. 
For this case the expected birth time delay was $\pm 247$ zs for $R=0.74$ \AA. 
Between these extreme cases the birth time delay showed a linear dependence on $\cos(\beta)$.
For comparison to the experimental data, the blue line in Fig. 3 C resembles this simple model.
Note that the model agreed with the prediction from Ref. \cite{Yudin2006} [Eq.~12], if one neglects the ionization potential.

Second, the red line shows the result from a more refined model.
It accounted for the fact that, other than in the optical double slit, in photoionization the two interfering waves are not simply spherical.
At the high photon energy used here, the atomic nondipole effect tilts the electron angular distributions from each center slightly in forward direction with respect to the photon propagation \cite{Grundmann2020b}.
This fact led to an additional angular shift of the interference pattern that slightly increased the slope of the red line as compared to the blue one.
The red line obtained from considering molecular photoionization in the time domain was in line with the prediction of calculations of molecular photoionization in the frequency domain if nondipole effects are included in full \cite{Chelkowski2018}.
This model is in reasonable agreement with the experimental results,
but more theoretical work including electron-electron correlation is needed to clarify the deviation.

In conclusion, we have shown that the birth of a photoelectron wave from a molecular orbital did not occur simultaneously across the molecule.
With an electron interferometric technique, we have observed the resulting birth time delay which was imprinted as a phase difference between the parts of the wave launching from the two sides of the hydrogen molecule.
The observed effect is general and does not only alter molecular photoionization but is also expected to be relevant for electron emission from solids and liquids.\\
The analogy between electron emission from the H$_2$ molecule and a classical double-slit experiment suggests that the birth time delay could be interpreted as the travel time of the photon from one molecular center to the other which is up to 247 zs for the average bond length of H$_2$.
Our experimental results support this picture, but studies targeting more complex molecules and applying more sophisticated theoretical models are necessary to further unveil the scope of birth time delay.
This work can function as a benchmark for such studies.

\newpage
\begin{figure}[h]
\centering
\includegraphics[width=0.9\textwidth]{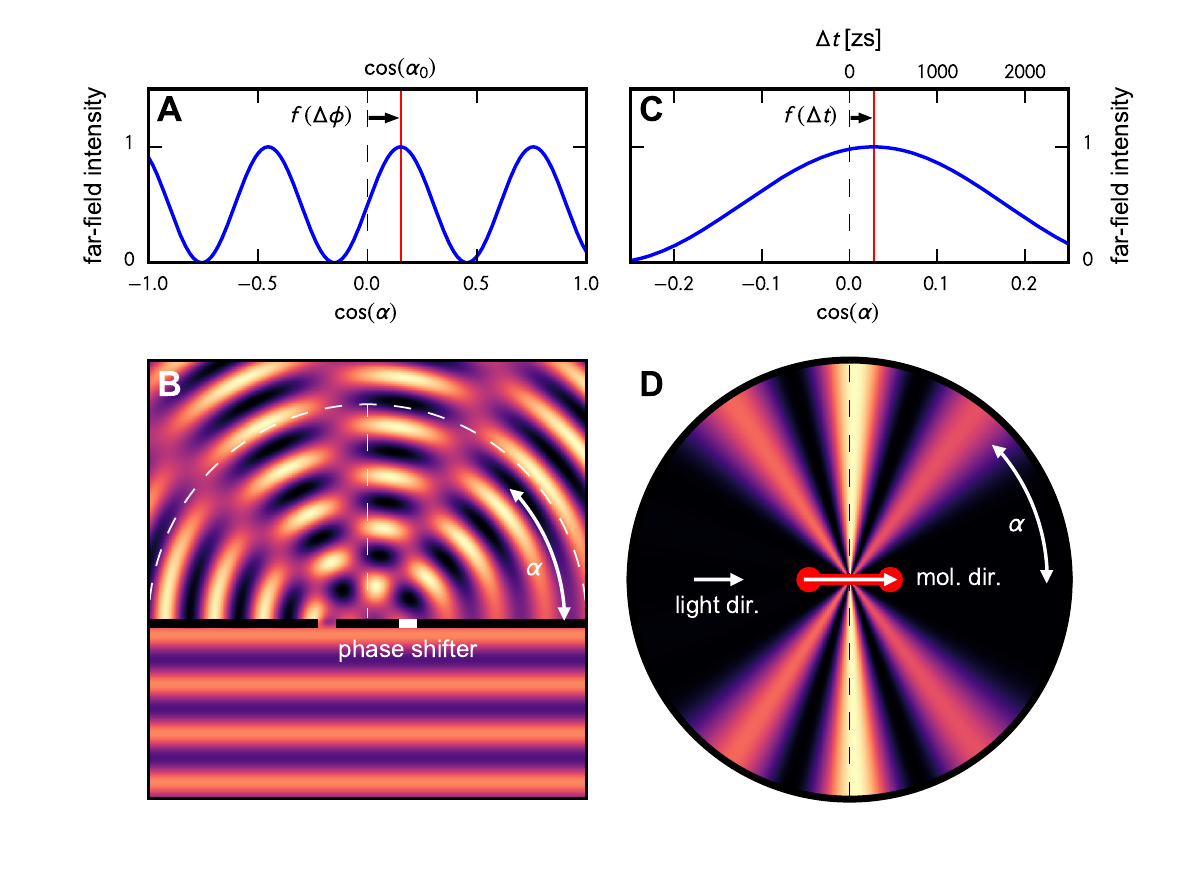}
\label{Fig1}
\end{figure}
\begin{singlespace}
\noindent {\bf Fig. 1.} Concept of birth time delay measurement. (A) Intensity distribution on a screen in the far field behind the double slit in panel B. (B) A plane wave impinges on a double slit. The phase shift [$\Delta \phi$] in the right slit causes a tilt of the interference pattern.
(C,D) Emission of a photoelectron wave from two indistinguishable atoms of a homonuclear diatomic molecule mimics the double-slit setup in panel B.
Here, the angle $\alpha$ is enclosed by the electron momentum vector and the molecular axis.
A time delay [$\Delta t$] between the emission from one of the two centers, e.g., originating from the travel time of the photon impinging from the left side in panel D, leads to a shift of the interference fringes in panel C. 
The ratio of slit distance [molecular bond length $R$, respectively] to wavelength is 1.65 in both cases (B,D).
In panel B the right-hand part of the wave is delayed by $\Delta \phi = \pi /2$, whereas in panel D a birth time delay of 247 zs causes $\Delta \phi \approx \pi/11$ for $R=0.74$ \AA.
\end{singlespace}

\newpage
\begin{figure}[h]
\centering
\includegraphics[width=0.55\textwidth]{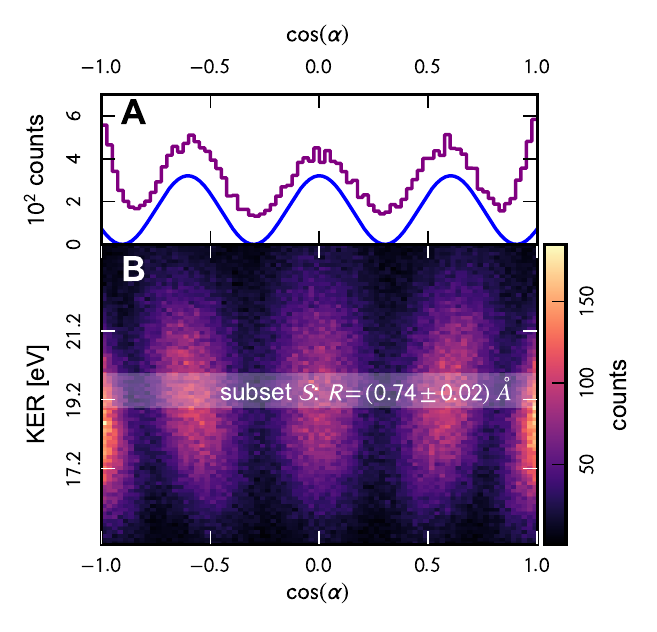}
\label{Fig2}
\end{figure}
\begin{singlespace}
\noindent {\bf Fig. 2.} Interference pattern of fast electrons [$E_e$ = (735$\pm$15) eV] from one-photon double ionization of H$_2$ by 800 eV circularly polarized photons for the average internuclear distance of {$R=(0.74\pm0.02)$~\AA} ~[purple line] in panel A and as function of $R$ in panel B. The blue line in panel A models a double-slit interference pattern for a slit distance $R$ = 0.74~\AA ~and $\lambda$ = 0.45~\AA,
which is the average de~Broglie wavelength of the fast electron. The subset $\mathcal{S}$ of the data is used for panel A and for the subsequent analysis of the birth time delay. 
\end{singlespace}

\newpage
\begin{figure}[h]
\centering
\includegraphics[width=0.9\textwidth]{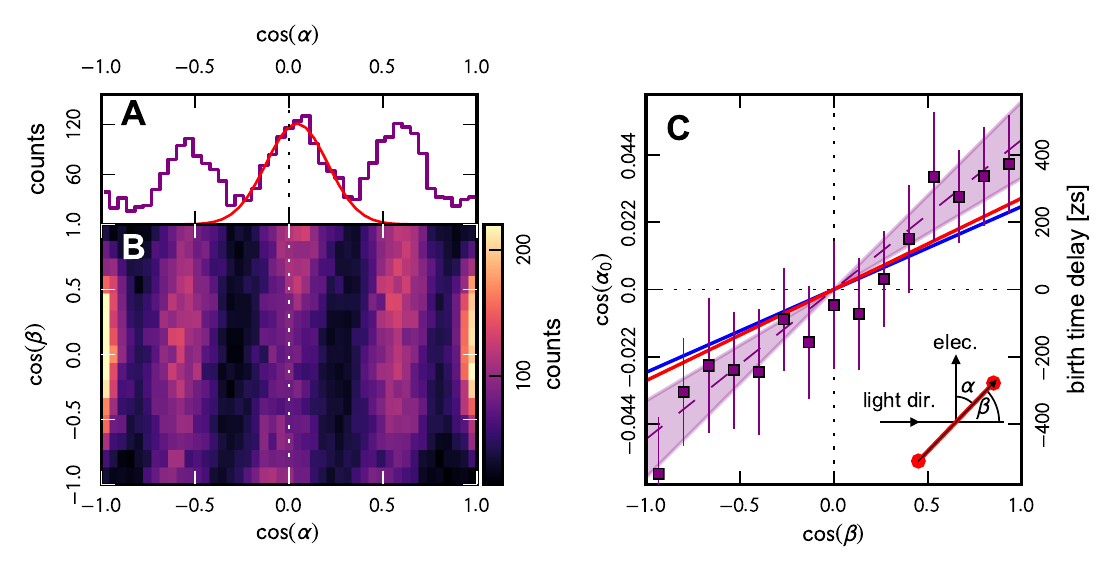}
\label{Fig3}
\end{figure}
\begin{singlespace}
\noindent {\bf Fig. 3.} Birth time delay of fast electrons [$E_e=($735$\pm$15) eV] from one-photon double ionization of H$_2$ by 800 eV circularly polarized photons for the average internuclear distance of {$R=(0.74\pm0.02)$~\AA} ~[selected subset $\mathcal{S}$ as shown in Fig.~2 B].
(A) Electron angular distribution with respect to the molecular axis which is aligned parallel to the light propagation direction [$\cos{(\beta)}>0.87$ corresponding to the top row of bins in panel B].
Red curve: Gaussian fit used to obtain the angular position of the zeroth-order maximum $\cos(\alpha_0)$.
(B) Electron angular distribution in the molecular frame of reference as function of $\cos(\beta)$.
Dashed line: perpendicular to molecular axis, i.e., location of the zeroth-order maximum in the absence of birth time delays.
(C) Location of the maxima of the zeroth-order interference fringe as function of $\cos(\beta)$. The maxima are obtained using Gaussian fits as indicated by the red line in panel A.
The error bars include statistical and systematic errors and the purple-shaded error range indicates the systematical error [see SM for further details].
Left axis: $\cos{(\alpha_0)}$, right axis: birth time delay calculated using Eq. \ref{equntau}.
The blue line resembles a birth time delay given by the travel time of light across the molecule [Eq.~3].
Red line: prediction combining atomic nondipole effects and the travel time of the photon [see text].
\end{singlespace}

\newpage
\section*{Acknowledgments}
We acknowledge DESY (Hamburg, Germany), a member of the Helmholtz Association HGF, for the provision of experimental facilities.
Parts of this research were carried out at PETRA~III and we would like to thank Jörn Seltmann and Kai Bagschik for excellent support during the beam time.
S.G. is very grateful for discussions on the topic with Anatoli~Kheifets and Heiner~Kremer.
{\bf Funding:} We acknowledge support by BMBF and by DFG.
K.F. acknowledges support by the German National Merit Foundation.
{\bf Author contributions:} M.S.S., T.J., K.F., N.S., A.P., L.K., S.E., M.W., S.G., D.T., L.Ph.H.S., R.D., and F.T. designed, prepared, and performed the experiment.
S.G. performed the data analysis.
S.G. and M.K. created the figures.
All authors contributed to the manuscript.
{\bf Competing interests:} None declared.
{\bf Data availability:} Data presented in this study are available on Zenodo \cite{DataGrundmann2020}.

\newpage    
\section*{Supplementary Materials}
\subsection*{Materials and Methods}
A Cold Target Recoil Ion Momentum Spectroscopy (COLTRIMS) reaction microscope (\textit{14}) has been employed to measure ion-ion-electron coincidences (see Fig. S1 for an illustration).
We intersected a supersonic gas jet of randomly oriented H$_2$ molecules with a synchrotron beam of 800~eV left-handed circularly polarized photons at right angle.
Charged reaction fragments were guided by weak electric (38.6~V/cm) and magnetic (36~G) fields from the interaction region towards two time- and position-sensitive detectors with an active area of 80 mm diameter (\textit{24,25}).
The spectrometer's ion arm had a length of 310 mm that was divided into an acceleration (120 mm) and a drift region (190 mm).
An electrostatic lens was created between acceleration and drift region in order to increase the momentum resolution of the ions in the detector plane.
The electron arm consisted of a single acceleration region of 73 mm length.
The magnetic field strength was chosen in order to facilitate 4$\pi$ solid angle detection for electrons with a kinetic energy of up to 420~eV.
Thus, independent of how the excess energy was partitioned, one electron was always detected.
From the flight times and the positions of impact, the initial momentum vector of each detected particle was deduced.
The missing electron's momentum vector was calculated using momentum conservation including the photon momentum.

To cover the full range of $-1<\cos(\beta)<1$, we label the two detected protons by their sequence of impact on the detector.
Thus, the molecule direction as depicted in Fig. 1 D and Fig. 3 C points from the proton detected first to the one detected second.

The experiment has been performed at the soft X-ray beamline P04 (\textit{26}) of the PETRA III electron storage ring (DESY, Hamburg, Germany) during timing mode (40 bunches, 192~ns bunch separation).
In order to increase the photon flux to an estimated maximum of {$1.6\times10^{14}$}~photons/s, we used a so-called pink beam by setting the monochromator to zeroth order.
Additionally, an aluminium blank mirror was used instead of the usual monochromator gratings of beamline P04.
To exclude low-energy photons, a foil filter was inserted into the beam path.
When using the pink beam of beamline P04, the bandwidth of the 800 eV beam is roughly 11 eV (resolving power $\approx$1/72 due to 72 undulator periods).
The foil filter does not affect the bandwidth.

The roughly 300.000 ion-ion-electron coincidences displayed in Fig.~2 have been obtained during a total acquisition time of approximately 132 h.

The accuracy to which we could specify the zero point of the momentum distribution of the fast electron was approximately $\pm$0.04 atomic units in all three directions in space (which resembles roughly 20\% of the magnitude of the photon momentum).
Within this systematic uncertainty, we determined the maximum and minimum slope of the linear fit function in Fig.~3~C (including $\pm 1 \sigma$ estimation error) and indicated this range by the purple-shaded area.
The statistical errors for each $\cos(\alpha_0)$ data point included in the error bars of Fig.~3~C are the standard deviations of the mean values from the Gaussian fits estimated as the square root of the respective diagonal element of the covariance matrix.

\subsection*{Supplementary Text}
We used left-handed circularly polarized photons because beamline P04 is currently not able to generate linearly polarized light due to technical reasons.
We do not expect any polarization dependence of the measured birth time delay.
In particular, the circular polarization does not add an additional phase shift beyond the birth time delay in the chosen body-fixed frame.
A circularly polarized light pulse can be imagined as a spiral that propagates through space.
In the moving frame of reference, this spiral itself does not rotate.
Only an observer (located on a stationary plane perpendicular to the propagation direction) that is passed by the spiral sees a rotating electric field vector.
In a reference plane that moves along with the light at phase velocity, as chosen in our description of the photoionization process, the electric field vector has the same length and points in the same direction at any time.
In this view, the physical interpretation underlying the birth time delay solely stems from the wavefront that sweeps across the molecule (see Eq. 3), which responds at different times accordingly.
The electric field vector associated with this wavefront has a fixed orientation and the different parts of the molecule are illuminated by the travelling wavefront with the same electric field vector properties at different times.

The contribution from circular dichroism on the phase of the electron interference pattern – as shown, e.g., in (\textit{11}) for photoionization of H2 at 160 eV photon energy – is very small at the much higher photon energy used here. Its effect has been averaged out by integrating over the azimuthal photoelectron angle with respect to the molecular-frame (i.e. in the plane perpendicular to $\cos(\alpha)$).
This way, a possible circular dichroism slightly broadens the peak but does not shift the mean angle like the phase shift due to the measured birth time delay.

\newpage
\begin{figure}[h]
\centering
\includegraphics[width=1.0\textwidth]{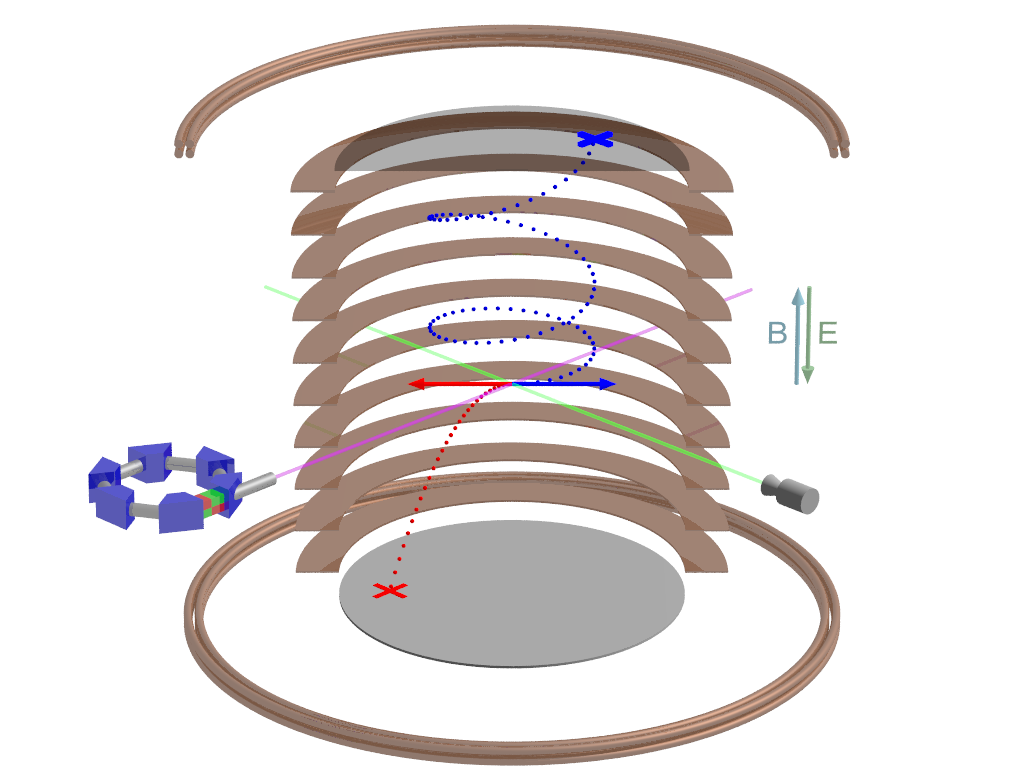}
\end{figure}
\begin{singlespace}
\noindent {\bf Fig. S1.} \textcolor{black}{Concept of cold target recoil ion momentum spectroscopy (COLTRIMS). A supersonic jet (green) of a target gas is crossed with synchrotron light (violet) at right angle. A homogeneous electric field $E$, generated by a spectrometer (copper plates), and a homogeneous magnetic field $B$, created by a Helmholtz coil pair (copper rings), guide the charged reaction fragments (red trajectory: ion, blue trajectory: electron) towards time- and position-sensitive detectors. The initial three-dimensional vector momentum of the reaction fragments (blue and red arrows) is calculated from the time-of-flight and position-of-impact on the detectors (marked with a red and a blue cross)}.
\end{singlespace}

\newpage
\begin{figure}[h]
\centering
\includegraphics[width=0.6\textwidth]{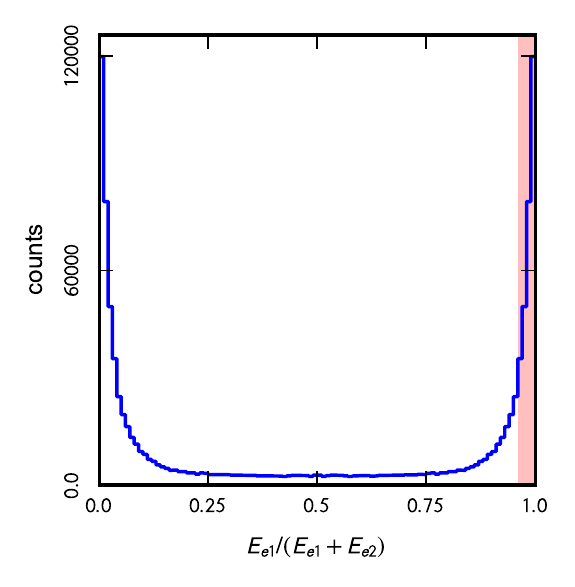}
\end{figure}
\begin{singlespace}
\noindent {\bf Fig. S2.} \textcolor{black}{Measured energy spectrum of one electron
from one-photon double ionization of H$_2$ by 800 eV circularly polarized photons.
If one electron carries most of the available energy, double-slit interference effects appear on a single-particle level even when the second electron is not observed.
Therefore, we restricted our investigation throughout this work to fast electrons that carried more than 96\% of the excess energy (red-shaded area).}
\end{singlespace}

\newpage
\begin{figure}[h]
\centering
\includegraphics[width=1.0\textwidth]{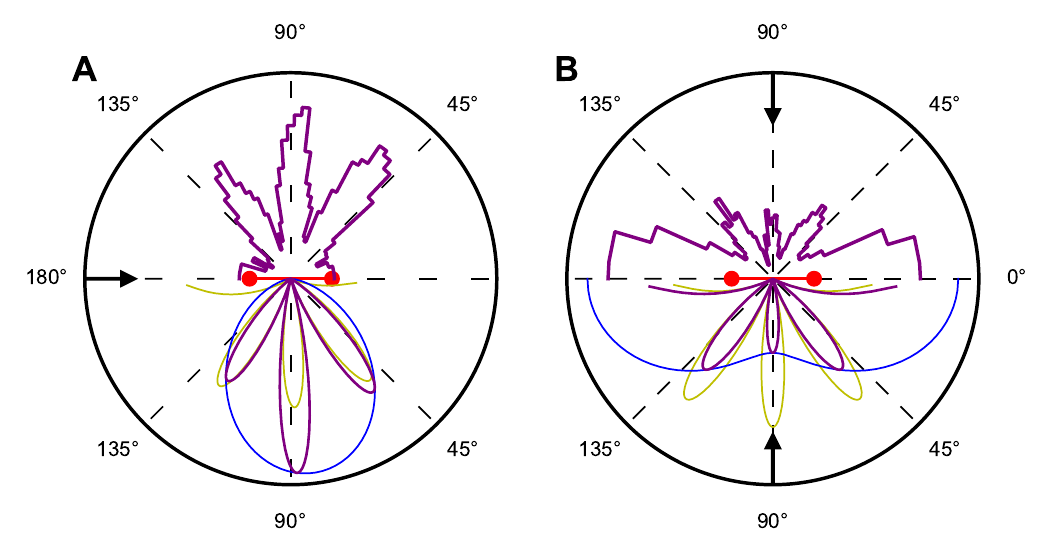}
\end{figure}
\begin{singlespace}
\noindent {\bf Fig. S3.} \textcolor{black}{Upper hemicircles: Measured molecular-frame photoelectron angular distributions of fast electrons [$E_e$ = (735$\pm$15) eV] from one-photon double ionization of H$_2$ by 800 eV circularly polarized photons for the average internuclear distance of $R$ = (0.74$\pm$0.02) \AA~ and different orientations between molecular axis and light propagation direction:
(A) Parallel alignment [$\cos(\beta)>0.87$] between light propagation (black arrow) and molecule (red barbell).
This is a different representation of the same data as in Fig. 3 A.
(B) Perpendicular alignment between light propagation and molecular axis [$\cos(\beta)=0\pm 0.065$].
Lower hemicircles, yellow lines: Double-slit interference patterns for $R=0.74$ \AA~and $\lambda=0.45$ \AA~(average de~Broglie wavelength of the fast electron), modified by the increased emission probability along the molecular axis [see Ref. (\textit{20})] and by the birth time delay.
Blue lines: 
Laboratory-frame photoelectron angular distributions transformed into the two-dimensional molecular frame of reference for the respective orientation of the molecule. Note that in panel B the light impinges from any direction perpendicular to the molecule.
Purple lines: Superposition of the modified double-slit interference pattern and the laboratory-frame envelope.
The resemblance between the model and the measured data is good in panels A and B, suggesting that the concept of two-center interference in photoionization of H$_2$ holds true independent of the molecular orientation.
The differences are most likely due to the integration over $\cos(\beta)$ necessary for the display of the experimental results.}\end{singlespace}

\end{document}